\documentclass[prd,twocolumn,showpacs,showkeys,amsmath,amssymb]{revtex4}

\usepackage{psfrag}
\usepackage{graphicx}

\newcommand{\brak}[1]{\left(#1\right)}    
\newcommand{\edg}[1]{\left[#1\right]}     

\newcommand{\diag}[1]{{\rm diag}\left(#1\right)}

\newcommand{\abs}[1]{\left|#1\right|}     

\begin{document}

\title{On Gravitational Waves in Spacetimes with a \\ Nonvanishing 
Cosmological Constant}

\author{Joachim N\"af} 
\email{naef@physik.uzh.ch}

\author{Philippe Jetzer}

\author{Mauro Sereno}

\affiliation{Institut f\"{u}r Theoretische Physik, Universit\"{a}t 
Z\"{u}rich, Winterthurerstrasse 190, CH-8057 Z\"{u}rich, Switzerland.} 

\date{October 27, 2008}

\begin{abstract}
\noindent We study the effect of a cosmological constant $\Lambda$ on the 
propagation and detection of gravitational waves. To this purpose we 
investigate the linearised Einstein's equations with terms up to linear 
order in $\Lambda$ in a de Sitter and an anti--de Sitter background spacetime. 
In this framework the cosmological term does not induce changes 
in the polarization states of the waves, whereas the amplitude gets modified 
with terms depending on $\Lambda$. Moreover, if a source emits a periodic 
waveform, its periodicity as measured by a distant observer gets modified. 
These effects are, however, extremely tiny and thus well below the 
detectability by some twenty orders of magnitude within present gravitational 
wave detectors such as LIGO or future planned ones such as LISA. 

\end{abstract}

\pacs{04.30.-w}
\keywords{Gravitational waves}

\maketitle

\section{Introduction}

\noindent 
The discovery that the expansion of the Universe is accelerating \cite{PR}, 
which can be interpreted as due to a cosmological constant $\Lambda$, has 
triggered a lot of recent works with the aim to study how $\Lambda$ affects 
e. g. celestial mechanics and the motion of massive bodies. In principle the 
cosmological constant should take part in phenomena on every physical scale.
For instance, it has been studied which limits on $\Lambda$ can be put from 
solar system measurements, such as the effect on the perihelion precession 
of the solar systemplanets \cite{I,W,KHM,js,sj1,sj3,IO}. The cosmological 
constant could also influence gravitationallensing \cite{RI,S} and play a role 
in the gravitational equilibrium of large astrophysical structures \cite{BN}. 
A natural question which arises is how the cosmological term affects 
gravitational waves. Clearly, we expect such an effect to be very tiny, 
nonetheless we believe that it is worthwhile to investigate it given the 
ongoing efforts in upgrading or building gravitational wave observatories 
either Earth bounded or in space.  

\noindent In this paper we study gravitational waves in spacetimes with a 
nonvanishing cosmological constant $\Lambda$ in the framework of perturbation 
theory with respect to de Sitter (dS) and anti--de Sitter (AdS) metrics. 
There are few articles in the literature devoted to the question on how the 
cosmological constant affects gravitational waves. Some approaches consider 
exact solutions of the Einstein's equations with a cosmological term relying 
on the Kundt class of spacetimes, which admit a non-twisting and 
expansion--free null vectorfield \cite{ORR, BP1, BP2}. In \cite{BP1, BP2} 
these spacetimes are interpreted as plane gravitational waves with 
polarizations ``$+$'' and ``$\times$'' which propagate on dS and AdS 
backgrounds.

\noindent A perturbative approach different from ours can be found in 
\cite{LU}, where the Einstein equations with a cosmological term are 
linearised with respect to a Minkowski background metric. By choosing a 
particular non Hilbert gauge this leads then to a Klein--Gordon equation 
and thus to a nontrivial dispersion relation.

\noindent Furthermore, we refer to some works on the scalar wave 
equation in dS and Schwarzschild-dS spacetimes \cite{POL,DR,YG,BH}. These 
treatments are, however, not directly connected to the present work, since 
the equations resulting from the linearisation of Einstein's equations are 
coupled partial differential equations for six independent variables.

\noindent The outline of the paper is as follows: in Section II we derive the 
linearised Einstein equations with respect to a dS or AdS background, which 
are represented by some generalized Klein--Gordon equations. Since a closed 
exact solution is not evident, we examine in Section III a perturbation expansion 
of these equations up to linear order with respect to $\Lambda$. In Section IV 
we calculate the corresponding first order contributions to the amplitudes. 
The effects on directly measurable quantities are discussed in Section V.

\noindent For the details of the linearisation of the Einstein equations with 
respect to an arbitrary differentiable background metric we refer e. g. to the 
textbooks \cite{STR,LL} or the review \cite{FH}.

\noindent As far as notation is concerned: Greek letters denote spacetime 
indices and range from $0$ to $3$, whereas Latin letters denote space indices 
and range from $1$ to $3$. If not stated otherwise, we use geometrical units 
($c = 1$ and $G = 1$).

\section{Linearised Einstein's Equations with Cosmological Term}

\noindent Let $(M,g_{\mu\nu})$ be a 4-dimensional Lorentz manifold with metric 
$g_{\mu\nu}$ of signature $(+,-,-,-)$. Let $R_{\mu\nu}$, resp. $R$, denote the 
Ricci tensor, resp. scalar, of $g_{\mu\nu}$. Then the vacuum Einstein equations 
with cosmological term read
\begin{equation}\label{einstein}
R_{\mu\nu} - \brak{\frac{R}{2} - \Lambda}g_{\mu\nu} = 0.
\end{equation}
\noindent In what follows we consider a perturbed metric
\begin{equation}\label{perturbed}
g_{\mu\nu} = \tilde{g}_{\mu\nu} + h_{\mu\nu},
\end{equation}
where $\tilde{g}_{\mu\nu}$ is a static background metric and $h_{\mu\nu}$ is a 
non--static perturbation with $|h_{\mu\nu}| \ll |\tilde{g}_{\mu\nu}|$. Up to 
first order in $h$ the indices are uppered and lowered by 
$\tilde{g}_{\mu\nu}$. Indicating the unperturbed Riemann tensor by 
$\tilde{R}_{\mu\nu\lambda\rho}$ and consequently the Ricci tensor, resp. 
scalar, by 
$\tilde{R}_{\mu\nu} = \tilde{R}^{\lambda}_{\phantom{\lambda}\mu\nu\lambda}$, 
resp. $\tilde{R} = \tilde{R}^{\lambda}_{\phantom{\lambda}\lambda}$, we can 
write the expansion of equation (\ref{einstein}) up to linear order in 
$h_{\mu\nu}$ as
\begin{eqnarray}\label{exph}
\tilde{R}_{\mu\nu} & + & R_{\mu\nu}(h) - \left(\frac{\tilde{R}}{2} + 
\frac{R(h)}{2} - \Lambda\right)\tilde{g}_{\mu\nu} \\ 
& - & \left(\frac{\tilde{R}}{2} - \Lambda\right)h_{\mu\nu} + \mathcal{O}(h^2) 
= 0, \nonumber
\end{eqnarray}
where the linear contributions to the Ricci tensor \cite{STR} and Ricci scalar 
are
\begin{eqnarray}\label{linricci}
R_{\mu\nu}(h) & = & \frac{1}{2}
\Big(h^{\lambda}_{\phantom{\lambda}\mu;\nu;\lambda} + 
h^{\lambda}_{\phantom{\lambda}\nu;\mu;\lambda} \\ 
& & - h^{\phantom{\mu\nu};\lambda}_{\mu\nu\phantom{;\lambda};\lambda} - 
h^{\lambda}_{\phantom{\lambda}\lambda;\mu;\nu}\Big), \nonumber \\ 
R(h) & = & R^{\lambda}_{\phantom{\lambda}\lambda}(h)  - 
h^{\lambda\rho}\tilde{R}_{\lambda\rho}. \nonumber
\end{eqnarray}
The semicolon denotes the covariant derivative with respect to 
$\tilde{g}_{\mu\nu}$. The terms in (\ref{exph}) which are independent of 
$h_{\mu\nu}$ satisfy equation (\ref{einstein}) with $\tilde{g}_{\mu\nu}$, 
\begin{equation}\label{constant}
\tilde{R}_{\mu\nu} - \brak{\frac{\tilde{R}}{2} - 
\Lambda}\tilde{g}_{\mu\nu} = 0.
\end{equation}
The terms linear in $h_{\mu\nu}$ are determined by 
\begin{equation}\label{linear1}
R_{\mu\nu}(h) - \frac{R(h)}{2}\tilde{g}_{\mu\nu} - \left(\frac{\tilde{R}}{2} 
- \Lambda\right)h_{\mu\nu} = 0. 
\end{equation}
In order to see explicitly the Klein--Gordon character of (\ref{linear1}), 
we rewrite this equation using the expressions in equation (\ref{linricci}) 
and the trace-reversed quantity 
$\tilde{\gamma}_{\mu\nu} := h_{\mu\nu} - \frac{h}{2}\tilde{g}_{\mu\nu}$, 
$h := h^{\lambda}_{\phantom{\lambda}\lambda}$. We are then left with
\begin{eqnarray}\label{linear2}
\tilde{\gamma}^{\phantom{\mu\nu};\lambda}_{\mu\nu\phantom{;\lambda};\lambda} &
+ & 
\tilde{\gamma}^{\phantom{\lambda\mu};\lambda}_{\lambda\mu\phantom{\lambda};\nu
}
+ \tilde{\gamma}^{\phantom{\lambda\nu};\lambda}_{\lambda\nu\phantom{\lambda};\mu
}
\\ & + & 2\tilde{R}_{\lambda\mu\rho\nu}\tilde{\gamma}^{\lambda\rho} - 
\tilde{R}_{\lambda\mu}\tilde{\gamma}^{\phantom{\nu}\lambda}_{\nu} - 
\tilde{R}_{\lambda\nu}\tilde{\gamma}^{\phantom{\mu}\lambda}_{\mu} \nonumber \\ &
- & 
\tilde{R}_{\lambda\rho}\tilde{g}_{\mu\nu}\brak{\tilde{\gamma}^{\lambda\rho} - 
\frac{\tilde{\gamma}^{\sigma}_{\phantom{\sigma}\sigma}}{2} 
\tilde{g}^{\lambda\rho}} \nonumber \\ 
& + & 2\Lambda\brak{\tilde{\gamma}_{\mu\nu} - 
\frac{\tilde{\gamma}^{\lambda}_{\phantom{\lambda}\lambda}}{2} 
\tilde{g}_{\mu\nu}} = 0, \nonumber 
\end{eqnarray}
In contrast to the corresponding result for the Einstein's equations without 
cosmological term (where instead of eq.(\ref{linear1}) we have 
$R_{\mu\nu}(h) = 0$ \cite{STR}), the equations (\ref{linear2}) feature in 
addition the two last terms.

\noindent In order to analyse further the equations (\ref{constant}) and 
(\ref{linear2}) we fix the background as follows. It is well known that a
dS resp. AdS metric solves the equations (\ref{einstein}) exactly. For our
purposes it is thus the natural choice for the background. We note that at
this point a Schwarzschild--de Sitter solution might have been chosen as well.
We avoid this since we are interested in a region of spacetime which is far from
sources of gravitational radiation. We now choose an appropriate coordinate
system for the background spacetime $(M,\tilde{g}_{\mu\nu})$. Let $p: I \subset
\mathbb{R} \to M$ be the locus of an observer at rest \cite{STR} and let 
$\phi: M \to \mathbb{R}^4$, $m \mapsto (t,x,y,z)$ be a coordinate chart such 
that $\phi(p(t)) = (t,0,0,0)$. An exact solution of (\ref{constant}) in the 
chart $\phi$ is given by
\begin{eqnarray}\label{desitter}
\tilde{g}_{00} & = & \brak{\frac{1-\frac{\Lambda}{12}r^2}{1 + 
\frac{\Lambda}{12}r^2}}^2, \\
\tilde{g}_{ii} & = & \frac{-1}{\brak{1+\frac{\Lambda}{12}r^2}^2}, \nonumber \\ 
\tilde{g}_{ij} & = & 0, \quad i\neq j, \nonumber 
\end{eqnarray}
where $r:=\sqrt{x^2+y^2+z^2}$. The solution (\ref{desitter}) is valid inside the
null horizon $r^2 = 12/|\Lambda|$, which depends on the choice of the observer
$p$. The apparent spacelike nature of the normal to this surface is due to
the use of isotropic coordinates. For later use we denote the corresponding
hypersurface in our coordinate chart by $\Omega := \{(x,y,z) \in \mathbb{R}^3 \
| \ r^2 < 12/|\Lambda|\}$. However, for the following applications it
suffices to consider a region which is much smaller than $\Omega$. The metric
(\ref{desitter}) was first introduced in \cite{dS} and is therefore known as dS
resp. AdS metric according to $\Lambda > 0$ resp. $\Lambda < 0$. Its Riemann
tensor is given by
\begin{equation}\label{riemann}
\tilde{R}_{\mu\nu\lambda\rho} = \frac{\Lambda}{3}\brak{\tilde{g}_{\mu\lambda} 
\tilde{g}_{\nu\rho} - \tilde{g}_{\mu\rho}\tilde{g}_{\nu\lambda}}.
\end{equation}

\noindent The equations (\ref{linear2}) form a family of ten coupled 
generalised Klein--Gordon equations for which an algorithm providing closed 
solutions is not known. We point out that the high symmetry of dS
resp. AdS allows to derive exact solutions of equation (\ref{einstein})
\cite{BP1,BP2}. Moreover, if we impose the Hilbert gauge condition
$\tilde{\gamma}^{\phantom{\mu\nu};\nu}_{\mu\nu} = 0$, then the trace of
equation (\ref{linear2}) turns into a simple Klein--Gordon equation for the
trace of $\tilde{\gamma}_{\mu\nu}$ on dS resp. AdS space,
\begin{equation}\label{trace}
\tilde{\gamma}_{\phantom{\lambda};\lambda}^{;\lambda} + 2\Lambda\tilde{\gamma}
= 0,
\end{equation}
which may be solved exactly by using separation of variables \cite{POL}.
However, since we are interested in the physical consequences of the
cosmological constant for all the components $\tilde{\gamma}_{\mu\nu}$ (and not
just for the trace) in the regime of a metric perturbation, equation
(\ref{trace}) does not provide enough information. Moreover, the perturbed
solutions derived below are traceless, thus only the trivial solution of
(\ref{trace}) is relevant for our purposes.
 
\noindent Note that the contraction of the equations (\ref{linear2}) with the
stationary Killing field $(\partial_t)^{\lambda}$ might lead to simpler
equations. However, the resulting equations are still non--scalar, as this is
suggested by the equations (\ref{D1}) below. Thus the derivation of an analytic
result, if possible, would be quite involved.

\noindent Although it would be useful to supplement the perturbative
calculation below with analytic results in order to gain more confidence in the
former, it seems that the effort for such a program would exceed the derivation
of the perturbative results and definitely goes beyond the scope of the present
work. We therefore content ourselves with an expansion of (\ref{linear2}) with
respect to $\Lambda$ up to linear order.

\noindent We remark that such an expansion with respect to $\Lambda$ is 
consistent with the expansion with respect to $h_{\mu\nu}$. Equations 
(\ref{constant}) may also be expanded with respect to $\Lambda$, and the 
coefficients of each order fulfill the equations subsequently. This point of 
view would correspond to a one--parameter perturbation of the Minkowski metric 
$\eta_{\mu\nu} = \diag{1,-1,-1,-1}$ of the form 
\begin{equation}
g_{\mu\nu} = \eta_{\mu\nu} + \sum_{n=1}^{\infty}\Lambda^n h^{(n)}_{\mu\nu}.
\end{equation}
Each coefficient $h^{(n)}_{\mu\nu}$ could be written as
\begin{equation}\label{h}
h^{(n)}_{\mu\nu} = \tilde{h}^{(n)}_{\mu\nu} + \bar{h}^{(n)}_{\mu\nu},
\end{equation}
where
\begin{equation}
\sum_{n=1}^{\infty}\Lambda^n \tilde{h}^{(n)}_{\mu\nu} = 
\tilde{g}_{\mu\nu}-\eta_{\mu\nu} \quad \textrm{and} \quad 
\sum_{n=1}^{\infty}\Lambda^n \bar{h}^{(n)}_{\mu\nu} = h_{\mu\nu}. 
\end{equation}
In other words, the $\tilde{h}^{(n)}_{\mu\nu}$ contain the contributions from 
the background, whereas the $\bar{h}^{(n)}_{\mu\nu}$ describe the waveform. 
Since $\Lambda$ carries the physical unit of $(\mathrm{Length})^{-2}$, 
the $n^{\mathrm{th}}$ order coefficients of the expansions above carry the 
physical unit $(\mathrm{Length})^{2n}$. 

\section{Approximate Solution of the Linearised Equations}

\noindent In particular the perturbation expansion with respect to $\Lambda$ 
is based on the assumption $r \ll \sqrt{12/|\Lambda|}$. We collect terms 
proportional to $\Lambda^n$ and denote them by $\mathcal{O}(\Lambda^n)$. 
For $r \ll \sqrt{12/|\Lambda|}$ equation (\ref{desitter}) yields
\begin{equation}\label{explambda1}
\tilde{g}_{\mu\nu} = \eta_{\mu\nu} + \mathcal{O}(\Lambda).
\end{equation}
Then the connection coefficients are of order $\mathcal{O}(\Lambda)$ 
and so are $\tilde{R}_{\mu\nu\lambda\rho}$ and $\tilde{R}_{\mu\nu}$ by 
(\ref{riemann}), such that equation (\ref{linear2}) may be written as 
\begin{equation}\label{explambda2}
\gamma^{\phantom{\mu\nu},\lambda}_{\mu\nu\phantom{,\lambda},\lambda} +
\gamma^{\phantom{\lambda\mu},\lambda}_{\lambda\mu\phantom{\lambda},\nu} +
\gamma^{\phantom{\lambda\nu},\lambda}_{\lambda\nu\phantom{\lambda},\mu} +
\Lambda D_{\mu\nu}(\gamma) + \mathcal{O}(\Lambda^2) = 0,
\end{equation}
where $\gamma_{\mu\nu} := h_{\mu\nu} - \frac{h}{2}\eta_{\mu\nu}$, the comma 
denotes partial derivatives, the indices are uppered and lowered by 
$\eta_{\mu\nu}$, and $D_{\mu\nu}$ is a linear hyperbolic differential operator 
of second order. We are thus led to consider the variable $\gamma_{\mu\nu}$ 
instead of $\tilde{\gamma}_{\mu\nu}$. These variables differ if the trace $h$ 
does notvanish. However, the solutions which are considered in the following 
sections are trace--free, and therefore they satisfy 
$\tilde{\gamma}_{\mu\nu} = \gamma_{\mu\nu} = h_{\mu\nu}$.

\noindent Hereafter we will neglect the terms of order 
$\mathcal{O}(\Lambda^2)$. Then $\Lambda$ lends itself as expansion parameter 
for the following perturbation procedure.

\noindent Let $\square := \partial^{\lambda}\partial_{\lambda}$ denote the 
d'Alembert operator. We assume that the exact solution of the operator 
equation
\begin{equation}\label{pertop}
\square\gamma_{\mu\nu} +
\gamma^{\phantom{\lambda\mu},\lambda}_{\lambda\mu\phantom{\lambda},\nu} +
\gamma^{\phantom{\lambda\nu},\lambda}_{\lambda\nu\phantom{\lambda},\mu} +
\Lambda D_{\mu\nu}(\gamma) = 0
\end{equation}
can be written as a power series in $\Lambda$. Up to linear order we then have 
\begin{equation}\label{pertser}
\gamma_{\mu\nu} = \gamma_{\mu\nu}^{(0)} + \Lambda\gamma_{\mu\nu}^{(1)} + 
\mathcal{O}(\Lambda^2),
\end{equation}
where the coefficients $\gamma_{\mu\nu}^{(1)}$ carry the physical unit 
$(\mathrm{Length})^{2}$. Thus a comparison to the case of a vanishing 
cosmological constant is achieved simply by considering only the zeroth order 
terms. Plugging (\ref{pertser}) into (\ref{pertop}) yields by comparing
coefficients order by order
\begin{eqnarray}
\label{eqcoeff}
\square\gamma_{\mu\nu}^{(0)} +
\gamma^{(0)\phantom{,},\lambda}_{\lambda\mu\phantom{\lambda,},\nu} +
\gamma^{(0)\phantom{,},\lambda}_{\lambda\nu\phantom{\lambda,},\mu} & = &
0,
\\ \square\gamma_{\mu\nu}^{(1)} & = & -D_{\mu\nu}(\gamma^{(0)}). \nonumber
\end{eqnarray}
On the first equation in (\ref{eqcoeff}) we impose the hilbert gauge 
condition $\gamma^{\phantom{\mu\nu},\nu}_{\mu\nu\phantom{\nu}} = 0$ 
and afterwards choose the transverse traceless gauge. Thus the fundamental 
solutions are plane gravitational waves with the two linear polarization 
states ``$+$'' and ``$\times$''.The solutions of the second equation in 
(\ref{eqcoeff}) are then determined by the expression 
\begin{equation}\label{conv}
\gamma_{\mu\nu}^{(1)} = -\mathcal{G}*D_{\mu\nu}(\gamma^{(0)}),
\end{equation}
where
\begin{equation}
\mathcal{G}(t,r) = \frac{\delta(t-r)\theta(t)}{4\pi r}
\end{equation}
is the Green's function of the d'Alembert operator and the star denotes the 
convolution. The domain of integration in (\ref{conv}) is $\mathbb{R}^3$, 
which may be interpreted as lowest order approximation of $\Omega$. In order 
to avoid divergences we need to choose an appropriate decrease for the 
amplitude of $\gamma_{\mu\nu}^{(0)}$ for $r \to \infty$. For our case a power 
counting argument requires the asymptotic behaviour 
$\left|\gamma_{\mu\nu}^{(0)}\right| \sim r^{-\alpha}$ for $\alpha > 2$. 
Accordingly one requires the boundary conditions 
\begin{equation}
\lim_{r \to \infty}\gamma_{\mu\nu}^{(1)} = 0 \qquad \textrm{and} \qquad 
\lim_{r \to \infty}\gamma^{(1)}_{\mu\nu,\lambda} = 0. 
\end{equation}
However, in the following sections we will consider only a small region of 
spacetime, so that the question of the behaviour for $r \to \infty$ is not 
essential. We will, therefore, assume that the intersection of any spacelike 
hypersurface with the support of $\gamma_{\mu\nu}^{(0)}$ and thus the domain 
of integration in (\ref{conv}) is compact.

\noindent It is understood that in contrast to Minkowski space the notion of 
planarity of a wavefront has to be modified for waves in curved spacetime. 
In the framework of exact solutions of the Einstein's equations this is 
achieved by demanding that the spacetime admits a null vectorfield which is 
non-twisting and expansion-free. 

\noindent However, in the perturbative approach we naturally assume that the 
wave front is a hyperplane up to lowest order. In a consistent perturbation 
expansion we are thus advised to assume that the fundamental solutions 
$\gamma^{(0)}_{\mu\nu}$ of the first equation in (\ref{eqcoeff}) is a 
Minkowski--plane wave. As mentioned above we restrict the support of 
$\gamma^{(0)}_{\mu\nu}$. In doing so we need to avoid further destruction of 
the symmetries of plane waves. Therefore we choose the domain of integration 
in (\ref{conv}) to be spherically symmetric and indicate it by 
$\Omega_{\mathcal{R}} := \{(x,y,z) \in \mathbb{R}^3 \ | \ r < \mathcal{R}\}$. 

\section{Plane Wave Propagation}

\noindent We now choose the coordinate chart $\phi$ such that 
$\gamma^{(0)}_{\mu\nu}$ is a plane transverse traceless solution and the 
non--vanishing components are $\gamma^{(0)}_{11}$, 
$\gamma^{(0)}_{22} = -\gamma^{(0)}_{11}$ and $\gamma^{(0)}_{12}$. 
These components are functions of the retarded time $z - t$ and describe 
thus a plane wave propagating in $z$--direction. The non--vanishing components 
of $D_{\mu\nu}(\gamma^{(0)})$ are then given by
\begin{eqnarray}\label{D1}
D_{01}(\gamma^{(0)}) & = & \frac{7}{6}\brak{x\partial_t\gamma^{(0)}_{11} + 
y\partial_t\gamma^{(0)}_{12}}, \\ 
D_{02}(\gamma^{(0)}) & = & \frac{7}{6}\brak{x\partial_t\gamma^{(0)}_{12} - 
y\partial_t\gamma^{(0)}_{11}}, \nonumber \\ 
D_{11}(\gamma^{(0)}) & = & \frac{r^2}{6}\brak{2\partial_{tt}\gamma^{(0)}_{11} - \partial_{zz}\gamma^{(0)}_{11}} \nonumber \\ 
& & - \frac{z}{6}\partial_{z}\gamma^{(0)}_{11} + \frac{2}{3}\gamma^{(0)}_{11}, \nonumber \\ 
D_{22}(\gamma^{(0)}) & = & - D_{11}(\gamma^{(0)}), \nonumber \\ 
D_{12}(\gamma^{(0)}) & = & \frac{r^2}{6}\brak{2\partial_{tt}\gamma^{(0)}_{12} - \partial_{zz}\gamma^{(0)}_{12}} \nonumber \\ 
& & - \frac{z}{6}\partial_{z}\gamma^{(0)}_{12} + \frac{2}{3}\gamma^{(0)}_{12}, \nonumber \\
D_{13}(\gamma^{(0)}) & = & \frac{5}{6}\brak{x\partial_z\gamma^{(0)}_{11} + 
y\partial_z\gamma^{(0)}_{12}}, \nonumber \\
D_{23}(\gamma^{(0)}) & = & \frac{5}{6}\brak{x\partial_z\gamma^{(0)}_{12} - 
y\partial_z\gamma^{(0)}_{11}}, \nonumber 
\end{eqnarray}
We now restrict ourselves to the investigation of the contributions 
to the ``$+$''--mode of $\gamma^{(0)}$. An analogous result may be derived for 
the ``$\times$''--mode. We have $\gamma^{(0)}_{11} = f(z-t)$ and 
$\gamma^{(0)}_{12} = 0$, where $f: D\subset\mathbb{R} \to \mathbb{R}$ is an
arbitrary 
smooth function. Equations (\ref{D1}) yield
\begin{eqnarray}
D_{01}(\gamma^{(0)}) & = & -\frac{7x}{6}f'(z-t), \\ 
D_{02}(\gamma^{(0)}) & = & \frac{7y}{6}f'(z-t), \nonumber \\ 
D_{11}(\gamma^{(0)}) & = & \frac{r^2}{6}f''(z-t) - \frac{z}{6}f'(z-t) + 
\frac{2}{3}f(z-t), \nonumber \\ 
D_{22}(\gamma^{(0)}) & = & -D_{11}(\gamma^{(0)}), \nonumber \\ 
D_{12}(\gamma^{(0)}) & = & 0, \nonumber \\ 
D_{13}(\gamma^{(0)}) & = & \frac{5x}{6}f'(z-t), \nonumber \\ 
D_{23}(\gamma^{(0)}) & = & -\frac{5y}{6}f'(z-t). \nonumber 
\end{eqnarray}
Thus we are able to calculate the first order corrections by using formula 
(\ref{conv}), i. e.
\begin{equation}
\gamma^{(1)}_{\mu\nu}(t,\vec{x}) = -\frac{1}{4\pi}\int_{\Omega_{\mathcal{R}}} 
\frac{D_{\mu\nu}(\gamma^{(0)})\left(t-|\vec{x}-\vec{\xi}|,\vec{\xi}\right)}
{|\vec{x}-\vec{\xi}|} \, d^3\xi. 
\end{equation}
In particular, all components vanish except
\begin{eqnarray}\label{gamma1.1}
\gamma^{(1)}_{11}(t,\vec{x}) & = & \frac{1}{24\pi}\int_{\Omega_{\mathcal{R}}}
\Bigg[(\vec{\xi})^2 f''\left(\xi_3 - (t-|\vec{x}-\vec{\xi}|)\right) \nonumber 
\\ & & - \xi_3 f'\left(\xi_3 - (t-|\vec{x}-\vec{\xi}|)\right) \\ 
\nonumber & & + 4f\left(\xi_3 - (t-|\vec{x}-\vec{\xi}|)\right)\Bigg] 
\frac{d^3\xi}{|\vec{x}-\vec{\xi}|} \nonumber \\ 
\gamma^{(1)}_{22}(t,\vec{x}) & = & -\gamma^{(1)}_{11}(t,\vec{x}). \nonumber 
\end{eqnarray}
This result indicates that in contrast to the amplitude the polarization 
remains unchanged up to this order, thus preserving the quadrupole character 
of gravitational radiation. Though an evaluation of the equation
(\ref{gamma1.1}) in general can hardly be carried out analytically for arbitrary
events $(t,\vec{x})$, it still may be computed along the locus $p(t)$ of the
observer using spherical coordinates. We now introduce physical units. Let 
$\gamma_{11}(t,\vec{0}) = f(\omega t)$, where $\omega$ denotes a frequency, 
and let $c$ denote the speed of light. Then the non--vanishing components 
of the perturbation $h_{\mu\nu}$ in (\ref{perturbed}) are determined by 
\begin{eqnarray}\label{res}
h_{11}(t,\vec{0}) & = & \gamma_{11}(t,\vec{0}) \approx \brak{\gamma^{(0)}_{11} 
+ \Lambda\gamma^{(1)}_{11}}(t,\vec{0}) \\
& = & f(\omega t) + \frac{\Lambda}{24\pi}
\Bigg[\frac{\mathcal{R}^3\omega}{3c}f'(-\omega t) \nonumber \\ 
& & +\frac{\mathcal{R}^2}{2}
\brak{f(-\omega t)-f\left(\frac{2\mathcal{R}\omega}{c}-\omega t\right)} 
\nonumber \\ & & -\frac{\mathcal{R}c}{\omega}\brak{5f^{\uparrow}(-\omega t)-
f^{\uparrow}\left(\frac{2\mathcal{R}\omega}{c}-\omega t\right)} \nonumber
\\ & & - \frac{2c^2}{\omega^2}\brak{f^{\uparrow\uparrow}(-\omega t)-
f^{\uparrow\uparrow}\left(\frac{2\mathcal{R}\omega}{c}-\omega t\right)}\Bigg],
\nonumber 
\end{eqnarray}
where we have denoted the primitive of any function $g:D\subset\mathbb{R} \to
\mathbb{R}$ by
\begin{equation}
g^{\uparrow}(t) := \int^t g(t')dt'.
\end{equation}
Due to the parameter $\mathcal{R}$, the formula (\ref{res}) is not yet in a 
form which allows an immediate meaningful physical interpretation. A priori 
$\mathcal{R}$ is a positive real number which measures the dimension of the 
support of $\gamma^{(0)}_{\mu\nu}$ in Minkowski spacetime. A posteriori we 
gather from equation (\ref{res}) that the perturbation expansion is reasonable 
if
\begin{equation}\label{lim1}
\lim_{\Lambda \to 0}\frac{\Lambda\mathcal{R}^3\omega}{c} = 
\lim_{\Lambda \to 0}\Lambda\mathcal{R}^2 = 
\lim_{\Lambda \to 0}\frac{\Lambda\mathcal{R}c}{\omega} = \lim_{\Lambda \to
0}\frac{\Lambda c^2}{\omega^2} = 0.
\end{equation}
Thus we formally obtain $\Lambda$--dependent constraints on $\mathcal{R}$ and 
$\omega$. If we impose the the geometrical optics limit, 
$\mathcal{R} \gg c/\omega$, we have 
\begin{equation}\label{lim2}
\frac{\mathcal{R}^3\omega}{c} \gg \mathcal{R}^2 \gg
\frac{\mathcal{R}c}{\omega} \gg \frac{c^2}{\omega^2},
\end{equation}
so that all the limites in (\ref{lim1}) follow from the first 
one. In Section V we give more comments on the interpretation of 
$\mathcal{R}$. In particular we find that for our purposes we can assume 
$\mathcal{R} \ll 1/\sqrt{|\Lambda|}$. Since $\omega$ is a constant parameter, 
the limits in (\ref{lim1}) are fullfilled. The condition 
$\left|\Lambda\gamma^{(1)}_{11}\right| \ll \left|\gamma^{(0)}_{11}\right|$ 
yields then $\Lambda\mathcal{R}^3\omega/c \ll 1$, which gives an upper bound 
on $\omega$.

\noindent As an illustration of the above results we consider the example 
$f(\omega t) := \sin(\omega t)$. Due to the conditions in equation (\ref{lim2})
we can neglect the terms with coefficients proportional to $c^2/\omega^2$,
$\mathcal{R}c/\omega$ and $\mathcal{R}^2$, so that to 
leading order we get
\begin{eqnarray}\label{sin1}
h_{11}(t,\vec{0}) & \approx &  \sin(\omega t) + 
\frac{\Lambda\mathcal{R}^3\omega}{72\pi c}\cos(\omega t).
\end{eqnarray}
Since $\Lambda\mathcal{R}^3\omega/c \ll 1$, this can also be written as 
\begin{eqnarray}\label{sin2}
h_{11}(t,\vec{0}) & \approx &  \sin\left(\omega\left(t + 
\frac{\Lambda\mathcal{R}^3}{72\pi c} \right)\right).
\end{eqnarray}
For a periodic $\gamma^{(0)}_{\mu\nu}$, equation (\ref{sin1}) shows that the 
correction $\gamma^{(1)}_{\mu\nu}$ features a modified amplitude, whereas 
(\ref{sin2}) yields a modification of the frequency. In the following section 
we show that $\mathcal{R}$ depends on the proper time of the observer. In 
general the frequency therefore changes with varying time. 

\section{Effects on Measurable Quantities}

\noindent The coordinate data in the in this section corresponds to the lowest 
order approximation of the chart $\phi$, which represents a Minkowski 
background. Consider a source which starts to emit gravitational radiation 
at some event $(-t_0,\vec{x}_0)$ so that an observer at large distance 
$|\vec{x}_0| = t_0$ would start to perceive an approximately plane wave 
at the event $(0,\vec{0})$. Assume that the wave at this event had the shape 
of the function $f$ up to lowest order. Let the observer at $p(t)$ carry out a 
measurement during a time interval $[0,\tau]$, such that $\tau \ll t_0$. In 
addition to the wave $f$, the observer would measure increasing retarded 
contributions $\gamma^{(1)}_{\mu\nu}$ with increasing $\tau$. These 
contributions originate from a spherical region within $r \leq \tau$. For the 
present measurement we thus have $\mathcal{R} = \tau$. Reasonably we have 
$\tau \ll 1/\sqrt{|\Lambda|}$ and therefore $\mathcal{R} \ll 1/\sqrt{|\Lambda|}$. 
Using again physical units, for $\Lambda \approx 10^{-52}\mathrm{m}^{-2}$ this 
yields
\begin{equation}\label{tau}
\tau \ll 10^{18} \mathrm{s} \approx 10^{11} \mathrm{yr}.
\end{equation}
Let $\tau_{\mathrm{yr}}$ denote the length of the measurement in years, and 
let $c \approx 3 \cdot 10^8 \mathrm{m}/\mathrm{s}$. Then the condition 
\begin{equation}\label{con1}
\Lambda c^2 \tau^3 \omega \ll 1
\end{equation}
and the geometrical optics limit $\tau \gg 1/\omega$ yield the following 
constraints on $\omega$:
\begin{equation}\label{omega}
\frac{1}{\tau_{\mathrm{yr}}} \cdot 10^{-7} \mathrm{Hz} \ll \omega \ll 
\frac{1}{\tau_{\mathrm{yr}}^3} \cdot 10^{15} \mathrm{Hz}.
\end{equation}
The condition (\ref{tau}) implies a non--vanishing range for the parameter 
$\omega$ in (\ref{omega}). For $\tau$ ranging from a couple of minutes up 
to several thousands of years, the radiation emitted by typical sources of 
gravitational waves features frequencies in this range.

\noindent The measurement via the equation for geodesic deviation is carried 
out analogously to the case $\Lambda = 0$ (cf. \cite{STR}, e. g.). We have 
\begin{equation}\label{geo}
\frac{d^2n^i}{dt^2} = -R^i_{\phantom{i}00j}n^{j},
\end{equation}
where $\vec{n} = (n^1,n^2,n^3)$ is the separation vector between two 
neighbouring members of a congruence of timelike geodesics \cite{STR}. 
We expand the Riemann tensor with respect to the perturbation $h_{\mu\nu}$: 
\begin{equation}
R_{\mu\nu\lambda\rho} = \tilde{R}_{\mu\nu\lambda\rho} + 
R_{\mu\nu\lambda\rho}(h) + \mathcal{O}(h^2),
\end{equation}
where the linear contribution to the Riemann tensor is given by \cite{LL} 
\begin{eqnarray}\label{linriemann}
R^{\mu}_{\phantom{\mu}\nu\lambda\rho}(h) & = & 
\frac{1}{2}\Big(h^{\mu}_{\phantom{\mu}\nu;\rho;\lambda} + 
h^{\mu}_{\phantom{\mu}\rho;\nu;\lambda} - 
h^{\phantom{\nu\rho};\mu}_{\nu\rho\phantom{;\mu};\lambda} \\
& & - h^{\mu}_{\phantom{\mu}\nu;\lambda;\rho} - 
h^{\mu}_{\phantom{\mu}\lambda;\nu;\rho} + 
h^{\phantom{\nu\lambda};\mu}_{\nu\lambda\phantom{;\mu};\rho}\Big). \nonumber 
\end{eqnarray}
For any measurement it is always possible to configure the detector such that 
it is sensitive only to the ``$+$''--mode of the wave \cite{FH}. We assume 
that this is the case in the following paragraphs. Therefore, in the present 
case we consider only the following components of $R_{\mu\nu\lambda\rho}$: 
\begin{eqnarray}
\tilde{R}^i_{\phantom{i}00j} & = & -\frac{\Lambda}{3}\delta^i_j + 
\mathcal{O}(\Lambda^2) \\ 
R^1_{\phantom{1}001}(h) & = & -\frac{1}{2}\brak{\partial_{tt}\gamma^{(0)}_{11}
 + \Lambda\brak{\partial_{tt}\gamma^{(1)}_{11}
 + \frac{1}{3}\vec{x}\cdot\nabla\gamma^{(0)}_{11}}} \nonumber \\ 
& & + \mathcal{O}(\Lambda^2) = -R^2_{\phantom{2}002}(h). \nonumber
\end{eqnarray}
Along the locus of $p(t)$ the components of equation (\ref{geo}) thus read 
\begin{eqnarray}
\frac{d^2n^1}{dt^2} & = & \edg{\frac{1}{2}\frac{d^2\gamma^{(0)}_{11}}{dt^2} + 
\Lambda\brak{\frac{1}{3} + \frac{1}{2}\frac{d^2\gamma^{(1)}_{11}}{dt^2}}}n^1,\\
\frac{d^2n^2}{dt^2} & = & \edg{-\frac{1}{2}\frac{d^2\gamma^{(0)}_{11}}{dt^2} + 
\Lambda\brak{\frac{1}{3} - \frac{1}{2}\frac{d^2\gamma^{(1)}_{11}}{dt^2}}}n^2, 
\nonumber \\ \frac{d^2n^3}{dt^2} & = & \frac{\Lambda}{3}n^3. \nonumber 
\end{eqnarray}
Let $n^i(t) = n^i_{(0)} + \delta n^i(t)$ with 
$|\delta n^i(t)| \ll |n^i_{(0)}|$. We simplify the notation by setting 
$\gamma^{(i)}_{11}(t) \equiv \gamma^{(i)}_{11}(t,\vec{0})$. Since 
$\gamma^{(1)}_{11}(0) = \frac{d\gamma^{(1)}_{11}}{dt}(0) = 0$ we are then 
left with
\begin{eqnarray}
\frac{n^1(\tau)}{n^1_{(0)}} & \approx & 1 + \frac{\delta n^1(0)}{n^1_{(0)}} - 
\frac{1}{2}\gamma^{(0)}_{11}(0) \\ 
& & + \tau\brak{\frac{1}{n^1_{(0)}}\frac{d(\delta n^1)}{dt}(0) - 
\frac{1}{2}\frac{d\gamma^{(0)}_{11}}{dt}(0)} \nonumber \\ 
& & + \frac{1}{2}\gamma^{(0)}_{11}(\tau) + \Lambda\brak{\frac{\tau^2}{6} + 
\frac{1}{2}\gamma^{(1)}_{11}(\tau)}, \nonumber \\ 
\frac{n^2(\tau)}{n^2_{(0)}} & \approx & 1  + \frac{\delta n^2(0)}{n^2_{(0)}} + 
\frac{1}{2}\gamma^{(0)}_{11}(0) \nonumber \\
& & + \tau\brak{\frac{1}{n^2_{(0)}}\frac{d(\delta n^2)}{dt}(0) + 
\frac{1}{2}\frac{d\gamma^{(0)}_{11}}{dt}(0)} \nonumber \\ 
& & - \frac{1}{2}\gamma^{(0)}_{11}(\tau) + \Lambda\brak{\frac{\tau^2}{6} - 
\frac{1}{2}\gamma^{(1)}_{11}(\tau)}, \nonumber \\
\frac{n^3(\tau)}{n^3_{(0)}} & \approx & 1 + \frac{\delta n^3(0)}{n^3_{(0)}} + 
\frac{\tau}{n^3_{(0)}}\frac{d(\delta n^3)}{dt}(0) + \frac{\Lambda \tau^2}{6}. 
\nonumber
\end{eqnarray}
The contributions from the background thus induce an isotropic dilatation 
proportional to $\tau^2$ which reflects the expansion of the universe. These 
terms may also be derived from the coefficient $\tilde{h}_{\mu\nu}^{(1)}$ in 
equation (\ref{h}). From equation (\ref{res}) we deduce that for 
$\mathcal{R} = \tau$ the dominant term in $\gamma^{(1)}_{\mu\nu}(\tau)$ is 
proportional to $\tau^3$. In addition to a modification of the amplitude, for 
a periodic $\gamma^{(0)}_{\mu\nu}(\tau)$ this term leads to a loss of 
periodicity of the zeros of $\delta n^i(\tau)$. The term proportional to 
$\tau$ features the same consequences, whereas the term proportional to 
$\tau^2$ only affects the amplitude.

\begin{figure}
\resizebox{\hsize}{!}{\includegraphics{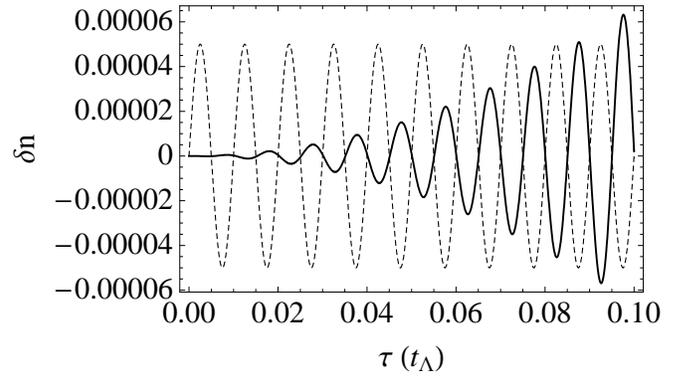}}
\caption{Magnified view of the contribution of $\Lambda$ to the geodesic 
separation. The bold line is the contribution due to $\Lambda$ coupled to 
the wave, whereas the dashed line is the unperturbed signal depleted by a 
factor $10^{-4}$. Obviously, $\Lambda$ affects in principle both amplitude 
and periodicity. While still preserving the ordering 
$1/\omega \ll \tau \ll t_\Lambda(\equiv 1/(c \sqrt{|\Lambda|}) 
\sim 10^{10}~\mathrm{years})$, 
we are considering not realistic time-scales for the wave form, i.e. a 
duration event $\Delta \tau = 10^{-1}t_\Lambda$ and a frequency 
$f = 10/\Delta \tau$.}
\label{delta_n_Lambda_GW}
\end{figure}

\begin{figure}
\resizebox{\hsize}{!}{\includegraphics{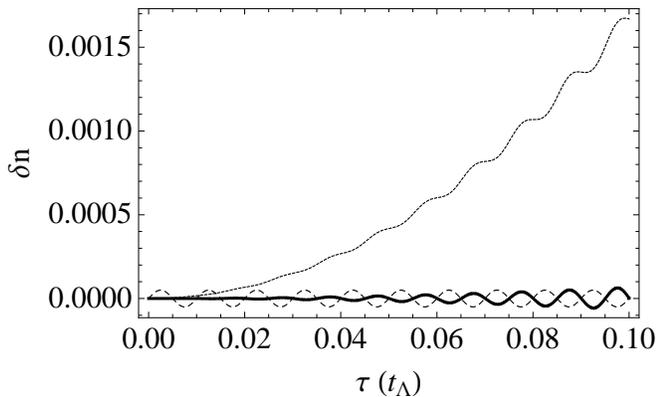}}
\caption{The same as fig. \ref{delta_n_Lambda_GW}, but including the dotted 
line which accounts for the isotropic expansion too.} 
\label{delta_n_Lambda}
\end{figure}

\noindent In the following example we again introduce physical units and 
illustrate the qualitative behaviour of $\delta n^1(\tau)$. Consider a 
source which starts to emit a wave at an event $(-c t_0,0, 0, z_0)$ with 
$c t_0 = |z_0|$ and $t_0 \gg \tau$. Let the source emit radiation during 
a time interval of length $s$. Moreover, assume that at the event 
$(0,\vec{0})$ the observer would perceive a sine wave up to lowest order. Then 
\begin{equation}
\gamma^{(0)}_{11}(\omega t) = \varphi(\omega t) := 
\left\{
\begin{array}{ll}
\sin(\omega t), & 0 \le t \le s\\
0, & \textrm{otherwise}. 
\end{array}
\right.
\end{equation}
We choose the initial conditions 
\begin{eqnarray}
\delta n^1(0) & = & \frac{n^1_{(0)}}{2}\gamma^{(0)}_{11}(0) \qquad \text{and}\\
\frac{d(\delta n^1)}{dt}(0) & = & 
\frac{n^1_{(0)}}{2}\frac{d\gamma^{(0)}_{11}}{dt}(0). \nonumber 
\end{eqnarray}
Equation (\ref{res}) with $\tau = \mathcal{R}$ and $f = \varphi$ then leads to
\begin{equation}
\begin{array}{rcl}
\delta n^1(\tau) & \approx & 
\displaystyle\frac{1}{2}\gamma^{(0)}_{11}(\omega\tau) + 
\Lambda\brak{\frac{c^2\tau^2}{6} + \frac{1}{2}\gamma^{(1)}_{11}(\omega\tau)} \\
& = & 
\left\{
\begin{array}{l}
\displaystyle\frac{1}{2}\sin(\omega\tau) + \frac{\Lambda}{24\pi}
\Bigg[c^2\tau^2\brak{4\pi-\frac{1}{2}\sin(\omega\tau)} \\ 
+ \displaystyle\frac{c^2\tau}{\omega}\cos(\omega\tau) +
\displaystyle\frac{2c^2}{\omega^2}\sin(\omega\tau)\Bigg], 
\quad 0 \le \tau \le s \\
\displaystyle\frac{\Lambda c^2\tau^2}{6}, \quad \textrm{otherwise}. 
\end{array}
\right.
\end{array}
\end{equation}
Figs. \ref{delta_n_Lambda_GW} and \ref{delta_n_Lambda} show the contribution 
of $\Lambda$ to the geodesic separation due to the wave. As shown in the plots 
$\Lambda$ affects both the amplitude and the frequency. In fig. 
\ref{delta_n_Lambda} the contribution from the isotropic expansion is also 
included. 

\begin{figure}
\resizebox{\hsize}{!}{\includegraphics{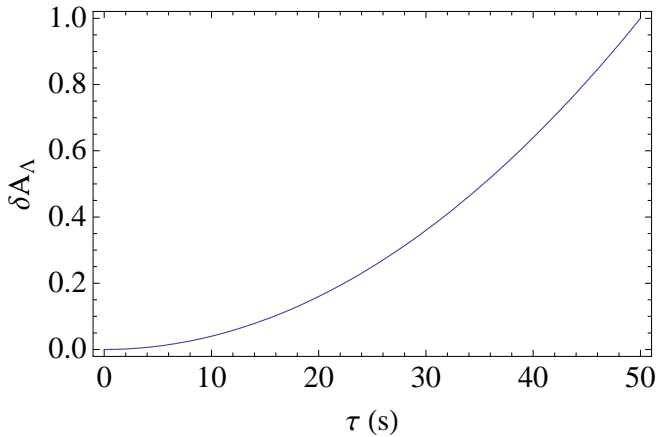}}
\caption{Contribution to the amplitude of the wave form due to 
$\Lambda$, $\delta A_\Lambda$, for a typical neutron star--neutron star 
inspiral in the LIGO band. We have considered 
$\Lambda = 10^{-52}\mathrm{m}^{-2}$, a frequency of 
$f(=\omega/ 2\pi) =  200~\mathrm{Hz}$ and a duration of $10^4$ cycles. The 
amplitude is normalized as to be unitary at the end of the detection, when 
$\delta A_\Lambda \simeq 3 \times 10^{-34}$; time units are in seconds.} 
\label{delta_A_Lambda}
\end{figure}

\begin{figure}
\resizebox{\hsize}{!}{\includegraphics{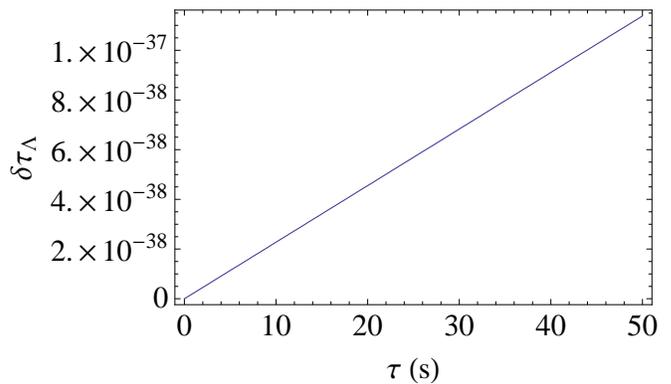}}
\caption{The same as fig.~\ref{delta_A_Lambda} for the phase shift 
$\delta \tau_\Lambda$.  The time unit on the $y$-axis is given by the 
unperturbed period, $T = 2 \pi/\omega 5\times 10^{-3}~\mathrm{s}$.} 
\label{delta_tau_Lambda}
\end{figure}

\noindent The shape of the amplitude as well as the approximate change of the 
frequency are explicitly apparent if we assume (\ref{con1}) and the 
geometrical optics limit and write the wave--dependent part for 
$0 \le \tau \le s$ in the form
\begin{eqnarray}\label{deltanwave}
\delta n_{\textrm{wave}}^1(\tau) = \frac{1}{2}\left(1-\delta A_\Lambda\right) 
\sin\left(\omega\left(\tau+\delta \tau_\Lambda\right)\right) 
\end{eqnarray}
with
\begin{eqnarray}
\delta A_\Lambda & = & \frac{\Lambda c^2 \tau^2}{24 \pi} \qquad \text{and}\\ 
\delta \tau_\Lambda & = & \frac{ 2 \Lambda c^2 \tau }{\omega^2}. \nonumber 
\end{eqnarray}
The functions $\delta A_\Lambda$ resp. $\delta \tau_\Lambda$ are shown in fig. 
\ref{delta_A_Lambda} resp. fig. \ref{delta_tau_Lambda} for a typical neutron 
star--neutron star inspiral in the LIGO band. 

\noindent As seen in equation (\ref{deltanwave},) for a positive value of 
$\Lambda$ the amplitude decreases, which might be due to the expansion induced 
by $\Lambda$. Indeed, we expect that an accelerated expansion stretches the 
wave [...]

\noindent Let $\tau_{\mathrm{day}}$ denote the length of the measurement 
in days. Then the relative weight of the leading $\Lambda$--dependent term for 
this example is of order
\begin{equation}\label{quan1}
\abs{\frac{\Lambda c^2\tau^2}{48\pi}} \approx 5 \cdot 10^{-28} \cdot 
\tau_{\mathrm{day}}^2.
\end{equation}

\noindent If the amplitude of the wave does not vanish before the measurement 
starts, i. e. if the function $f(t)$ unlike $\varphi(t)$ does not vanish for 
$t < 0$, we gather from the general result (\ref{res}) that then the leading 
term proportional to $\tau^3$ is present. The relative weight of this term 
depends on $\omega$ and thus on the type of the source of radiation. We have 
\begin{equation}\label{quan2}
\abs{\frac{\Lambda}{24\pi}\cdot\frac{\omega c^2\tau^3}{3}} \approx 2.5 
\cdot 10^{-23} \cdot \omega_{\mathrm{Hz}}\tau_{\mathrm{day}}^3,
\end{equation}
where $\omega_{\mathrm{Hz}}$ measures the frequency in Hertz. For compact 
sources $\omega$ is related to the size and the mass of the source. The size 
is bounded below by the Schwarzschild radius of the mass. This yields an upper 
bound on the frequency given by $\omega \approx 10^4\mathrm{Hz}$ \cite{FH}. 
Equation (\ref{quan2}) then leads in the best case to
\begin{eqnarray}
\abs{\frac{\Lambda}{24\pi}\cdot\frac{\omega c^2\tau^3}{3}} \lesssim 2.5 \cdot 
10^{-19} \cdot \tau_{\mathrm{day}}^3.
\end{eqnarray}
In principle the effects of $\Lambda$ are measurable if the signal to noise 
ratio (SNR) of the detector is sufficiently large. Present as well as planned 
observatories however do not feature the required accuracy. For example the 
Earthbounded detector advanced LIGO achieves a $\textrm{SNR} \approx 10$ for 
the inspiral of compact objects of mass $m \approx 10^2M_{\odot}$ at a 
frequency $\omega \approx 10^2\mathrm{Hz}$ \cite{MB}. Then the detectability 
of the effects of $\Lambda$ may be measured by
\begin{eqnarray}
\textrm{SNR}_{\Lambda,\textrm{LIGO}} & \approx & 2.5 \cdot 10^{-23} \cdot 
\omega_{\mathrm{Hz}} \tau_{\mathrm{day}}^3 \textrm{SNR}_{0,\textrm{LIGO}} \\ 
& \approx & 2.5 \cdot 10^{-20}\tau_{\mathrm{day}}^3. \nonumber
\end{eqnarray}
The planned spacebased observatory LISA on the other hand is expected to reach 
a $\textrm{SNR} \approx 10^4$ for the inspiral of supermassive black holes 
with $m \approx 10^6M_{\odot}$ at a frequency 
$\omega \approx 10^{-2}\mathrm{Hz}$ \cite{MB}. This yields
\begin{eqnarray}
\textrm{SNR}_{\Lambda,\textrm{LISA}} & \approx & 2.5 \cdot 10^{-23} \cdot 
\omega_{\mathrm{Hz}} \tau_{\mathrm{day}}^3 \textrm{SNR}_{0,\textrm{LISA}} \\ 
& \approx & 2.5 \cdot 10^{-21}\tau_{\mathrm{day}}^3. \nonumber
\end{eqnarray}
The corresponding SNR for the example with $f = \varphi$ can be calculated by 
considering (\ref{quan1}) instead of (\ref{quan2}). Then
\begin{eqnarray}
\textrm{SNR}_{\Lambda,\textrm{LIGO}} & \approx & 5 \cdot 10^{-28} \cdot 
\tau_{\mathrm{day}}^2 \textrm{SNR}_{0,\textrm{LIGO}} \\
& \approx & 5 \cdot 10^{-27}\tau_{\mathrm{day}}^2. \nonumber
\end{eqnarray}
respective
\begin{eqnarray}
\textrm{SNR}_{\Lambda,\textrm{LISA}} & \approx & 5 \cdot 10^{-28} \cdot 
\tau_{\mathrm{day}}^2 \textrm{SNR}_{0,\textrm{LISA}} \\
& \approx & 5 \cdot 10^{-24}\tau_{\mathrm{day}}^2. \nonumber
\end{eqnarray}
For $\tau_{\mathrm{day}} = 1$ e. g., the aforesaid observatories would have to 
increase their accuracy by at least twenty orders of magnitude in order to 
detect the effects of $\Lambda$ on the waveform radiated by the inspirals 
mentioned above. Thus even for a long but realistic period of measurement it 
is not possible to detect the effects of $\Lambda$ within the existing 
technology.

\section{Conclusions}

\noindent We investigated the linearised Einstein's equations with a 
cosmological term and derived explicit expressions for the corrections to the 
plane gravitational waves up to linear order in $\Lambda$. The polarization 
states of a wave remain unchanged in the presence of the cosmological term. 
This conclusion is consistent with the result obtained in \cite{BP1, BP2}. The 
amplitude as well as the frequency (for periodic radiation) though are 
modified with increasing time. However, these effects are very tiny and thus 
not detectable by present or planned detectors. 

\noindent We point out that one can not rule out the possibility that 
nonlinear effects originating from terms proportional to 
$h_{\mu\nu,\lambda}h_{\rho\sigma,\tau}$ in an expansion (\ref{exph}) could 
lead to effects on the waveform similar in size as the ones due to the 
cosmological term. However, as discussed for instance in \cite{FH}, such a 
perturbation term can be split into a slowly varying piece, and a rapidly 
varying one. The latter one would induce modifications on a much shorter 
timescale than the contribution due to the cosmological constant as 
considered here, and should thus be easily discriminated. On the other hand the 
long timescale contribution would modify the background. However, its 
timedepence might be different from the one due to the cosmological constant 
and thus making it still possible to distinguish the various effects. A 
detailed analysis of effects due to quadratic terms in $h$ is certainly quite 
involved and beyond the scope of the present work.

\noindent A mentionable phenomenon is eventually the connection between the 
cosmological constant and the mass of the graviton. Mass terms characterize 
Klein--Gordon equations and are connected to the dispersion relation. We do 
not go further into this question and refer to \cite{F,NN,L}.

\begin{center}
{\bf Acknowledgments}
\end{center}

\noindent M.S. is supported by the Swiss national science Foundation and by 
the Tomalla Foundation. We thank the referee for some clarifying suggestions.

\end{document}